\documentstyle[aps,preprint]{revtex}
\tightenlines
\begin{document}
\draft

\title{
Non linear dynamics of vortices in superconductors 
with short coherence length.}

\author{F. Guinea \\}
\address{
	Instituto de Ciencia de Materiales. \\
	Consejo Superior de Investigaciones Cient{\'\i}ficas. \\
	Cantoblanco. E-28049 Madrid. Spain.}

\author{Yu. Pogorelov}
\address{
	Departamento de F{\'\i}sica de la Materia Condensada. \\
	Universidad Aut\'onoma de Madrid. \\
	E-28049 Madrid. Spain.}

\date{\today}

\maketitle

\abstract{
		In superconductors where the coherence length is comparable to 
the Fermi wavelength, the vortex viscosity depends on the
velocity of the vortex, leading to non linear equations of motion.
The trajectories of vortices driven by a. c. fields show a variety of
behaviors as function of frequency. Finite pertubations
give rise to very long lived transients. The relevance of these results 
to experiments in high-T$_{\rm c}$ superconductors is discussed.


\narrowtext
\section{Introduction.}
	In the copper oxide superconductors, the coherence length,
$\xi$, is similar to the Fermi wavelength, $k_F^{-1}$, and to the
lattice spacing, $a$. The separation of energy levels within vortex
cores, $\delta \epsilon$ can be comparable, or larger, than the
width of these levels (ultra clean limit) and the temperature.
Using standard parameters, we estimate that $\delta \epsilon \sim 40$K,
in agreement with experimental observations\cite{Karrai}. 
The quantization of levels should change drastically dissipation 
processes at temperatures lower that $\delta \epsilon$.
The standard theory of flux flow dissipation\cite{BS} breaks down,
as it is based on the existence of fast relaxation processes
within the vortex core. Note that the limit $T \ll \delta \epsilon$
is also different from what is usually defined
as the ultra clean limit,
in which $\hbar \tau \ll \delta \epsilon$, but
$\delta \epsilon  < T$\cite{JETP}.

	Simple arguments show that energy dissipation due to vortex
motion should be exponentially reduced at low velocities.
The vortex viscosity behaves as 
$\eta ( v ) \sim e^{- v_0 / v}$\cite{GP}, where $v_0$ is 
given by $v_0 \sim \delta \epsilon \xi / \hbar$.
The existence of this threshold velocity can be understood
very simply: it corresponds to the situation for which the
inverse of the vortex-impurity collision time, $\tau_{coll} \sim
\xi / v$, becomes comparable to $\delta \epsilon / \hbar$.
Taking $\xi \sim 10$\AA, we find $v_0 \sim 4$ Km/s.

	The presence of such a term in the energy dissipation of
moving complex quantum objects with internal levels 
(atoms, molecules) is rather common\cite{PME}. The existence of
a gap in the excitation spectrum of the system implies 
that energy dissipation is
exponentially suppressed at low temperatures or velocities.

	The resulting equations of motion for the vortices become
highly non linear. In the following, we study the dynamics of vortices,
in the pinned regime, driven by a. c. fields. Our aim is to contribute
to the understanding of recent experiments which probe the
response of vortices at low temperatures\cite{GTD,Parks}.
These experiments show a viscosity systematically lower
than the value predicted by the Bardeen-Stephen theory.
These experiments have been performed in the range of 
$10^6$ to $10^9$ Hz. The previous analysis gives
$\tau_{coll}^{-1} \sim 10^{12}$ Hz. The number of
collisions per cycle is large enough to justify the 
theory presented in\cite{GP}.

	A velocity dependent viscosity has also been analyzed
in\cite{scaling}. In this work, it arises from the influence of
the pinning centers on an (almost) freely moving vortex,
at relatively high temperatures.
The total viscosity is always finite, and its value is
comparable to the Bardeen-Stephen prediction.
We do not consider that regime in the following. 

\section{Equation of motion.}

	We consider pinned vortices, with a linear restoring 
force. There is some controversy on the magnitude of the
Magnus force. To keep the number of parameters to a minimum,
we assume that vortices are carried by the applied current,
and that Galilean invariance fully determines this contribution
to the motion\cite{Ao}. 
Finally, we describe the external probe
through the superfluid velocity,
with a given time dependence.

	Then, the equations of motion, in the plane perpendicular to
the vortex axis,  are:

\begin{eqnarray}
\eta ( v ) v_x &= & - k_p x + \alpha v_y \nonumber \\
\eta ( v ) v_y &= & - k_p y - \alpha [ v_x - v_s ( t ) ]
\label{em} 
\end{eqnarray}

and $\eta ( v ) = \eta_0 e^{- v_0 / v} $. We use units such that
$k_p = 1$, $\eta_0 = 1$ and $v_0 = 1$. The equations (\ref{em}) are 
determined by the ratio $\alpha / \eta_0$. 
The only difference with the standard theory\cite{GTD,Parks,scaling}
lies in the assumption of a velocity dependent
viscosity. In the presence of
a periodic supercurrent, $v_s ( t ) = v_s^0 \cos ( \omega t )$, we
need also to specify the dimensionless
values of $ \alpha \omega $ and $v_s^0 $.

We now solve numerically eqs. (\ref{em}). We consider first the case
of a periodic driving current. We choose $\alpha = 1$ and
$v_s^0 = 0.5$. The value of $\alpha$ is
proportional to $n_s$, the condensate density,
and $\eta_0 \propto n_{imp}$, the concentration of impurities\cite{GP}.
Hence, $\alpha / \eta_0 \sim n_s / n_{imp}$.
Finally, $v_0$ is a fraction of the bulk
depairing velocity, $v_0 / v_{dp} \sim \delta \epsilon / \Delta$,
so that $v_s^0$ should not be greater than $v_0$.

If we could take $\eta = 0$, the solution of eqs.
(\ref{em}) depends only on the dimensionless ratio
$k_p / ( \alpha \omega )$.

The dynamics of the vortex depend strongly on the value of the 
driving frequency. To a first approximation, we can neglect the
friction term in eqs. (\ref{em}). Then, the motion of the vortex is
purely inductive, and the Hall angle changes from 0 ($\omega \ll 1$)
to $\pi / 2$ ( $\omega \gg 1$ ). This picture is qualitatively 
correct, as seen in fig.(\ref{vortex}).  The unit of length
is $\eta_0 v_0 / k_p$. Taking typical experimental
values (at low temperatures) for $k_p \sim 2 \times 10^5$ N / m$^2$ and 
$\eta_0 \sim 10^{-6}$ N s / m$^2$
\cite{GTD,Parks}, our unit of length is
$\sim 40$\AA.

The vortex orbit, however,
contains many higher harmonics (see next section). At sufficiently
high frequencies, these higher harmonics dominate the trajectory,
and the role of the dissipation needs to be taken into account.

The corresponding velocities are plotted in fig.(\ref{vel}).
In order to interpret this result, we need to consider the
influence of the dissipation. Firstly, it limits the maximum
values of the vortex velocity, which never becomes
much greater than 1, even close to resonance.
Above the resonant frequency, the motion becomes highly irregular.
Note that the number of degrees of freedom in eqs.(\ref{em})
suffices to generate chaotic behavior in the presence of a periodic
driving force. We have checked that eqs.(\ref{em}) do, indeed, show
chaotic behavior, for $\alpha \ll \eta_0$ and high frequencies. 

We have also analyzed the response to a finite pulse, as in the
experiments reported in\cite{Parks}. We find very long lived transients
after the pulse has been switched off. This effect can be understood
analytically. In the absence of a driving
force, we can take the time derivative of eqs. (\ref{em}), to
obtain:

\begin{eqnarray}
\eta ( v ) \left( 1 + \frac{v_0}{v} \right)
( v_x \dot{v}_x + v_y \dot{v}_y ) &= &- k_p v^2 
+ \alpha ( \dot{v}_y v_x - \dot{v}_x v_y ) \nonumber \\
\eta ( v ) ( \dot{v}_x v_y - \dot{v}_y v_x ) &= &\alpha
( \dot{v}_x v_x + \dot{v}_y v_y )
\label{transient}
\end{eqnarray}

which gives:

\begin{equation}
\dot{v} = \frac{k_p v \eta ( v ) }{\alpha^2 + \eta^2 ( v )
\left( 1 + \frac{v_0}{v} \right)}
\end{equation}

At long times, $v \ll v_0$ and $\eta ( v ) \ll \alpha$. Then:

\begin{equation}
\dot{v} \approx - \frac{k_p v \eta_0}{\alpha^2} e^{- \frac{v_0}{v}}
\end{equation}

which, to logarithmic accuracy, yields:

\begin{equation}
\lim_{t \rightarrow \infty} v \sim
\frac{v_0}
{ \log \left( 
\frac{ k_p \eta_0 t}{\alpha^2} 
\right) }
\label{asympt}
\end{equation}

Our numerical results are consistent with this asymptotic behavior.
Note that,  if the viscosity was independent of velocity,
transients decay exponentially,
with a well defined relaxation time, $\tau^{-1} \propto
\eta / ( \eta^2 + \alpha^2 )$.

\section{Non linear resistivity.}

	From the vortex velocities, plotted in fig. (\ref{vel}), we 
infer the induced voltage, by means of the Josephson relation,
$V \sim \vec{v} \times \vec{B}$. Hence, $\rho_{xx} \propto
v_y$ and $\rho_{xy} \propto v_x$. In order to analyze numerically 
the generation of higher harmonics, solve eqs. (\ref{em}) in the
presence of a periodic current, and perform a Fourier transform.

	The linear resistivities are plotted in fig. (\ref{resl}).
The two components which are finite in the limit of vanishing
dissipation, $\rm{Im} \rho_{xx}$ and $\rm{Re} \rho_{xy}$ are approximately
well described by the standard solution of eqs. (\ref{em}). 
We find a sharp resonance at $\omega \sim k_p  / \alpha$.
The resonance, however, is very  asymmetric, with a sharp rise at
low frequencies, and a slow decay at high frequencies.

The other components of the resistivity, $\rm{Re} \rho_{xx}$
and $\rm{Im} \rho_{xy}$ display more unusual behavior.
Both are very small at low frequencies. They become comparable to
the other components near the resonance frequency, and slowly decrease
at high frequencies. Note that, if the effective viscosity
remains finite as $\omega \rightarrow \infty$, these functions 
should tend to a constant. 
At low frequencies, we find the scaling law $\rho_{xy} (
\omega ) \sim
\rho_{xx}^2 ( \omega )$\cite{scaling} {\it as function of frequency}.
The effective viscosity near the resonance can be understood
from the standard, linear, solution to eqs. (\ref{em}).
Near the resonance, the vortex velocity behaves as
$v \sim \alpha v^0_s / \eta ( v )$. As $v$ cannot exceed 1, we find
that the value of $\eta ( v )$ at resonance should be similar to
the value of $\alpha$.

If the results shown in fig.(\ref{resl}) were analyzed in terms of
an effective, frequency dependent viscosity, we conclude that
$\eta_{eff} \sim 0$ for $\omega \ll k_p / \alpha$, $\eta_{eff} \sim
\rm{min} ( \alpha , \eta_0 )$ at resonance, $\omega \sim k_p
/ \alpha$, and that $\eta \rightarrow 0$ at high frequencies.

The resistivities also show a significant number of higher harmonics.
At low frequencies,
this effect is more pronounced in $\rm{Re} \rho_{xx}$ and $\rm{Im}
\rho_{xy}$, where the standard analysis indicates
that they vanish as $\eta \rightarrow 0$. It is unclear to us the
contribution to this effect of the long lived transients discussed
in the preceding section. The value of these harmonics 
is shown in fig (\ref{resh}) (note that the scales are different from
those in fig (\ref{resl}). The harmonics show also a maximum near the
resonance frequency. Note that the third harmonic is greater than the
second. In general, we find the odd harmonics to be larger that the
even ones.

\section{Conclusions.}

We have studied the motion of pinned vortices in the limit where the
spacing between quasiparticle states within the core is
greater than the inverse scattering time and the temperature.
In this regime, the conventional analysis of flux flow dissipation
ceases to be valid. The dependence of the energy dissipation on
the vortex velocity is highly non linear. There is a crossover velocity,
above which dissipation can be regarded as ^^ ^^ conventional ", and below 
which dissipation is exponentially suppressed.

The non linear effects dominate the behavior of $\rm{Re} \rho_{xx}$
and $\rm{Im} \rho_{xy}$ at low frequencies.
There is an asymmetric resonance
around $\omega_{res} \approx k_p / \alpha$, where $k_p$ is the pinning constant, 
$\alpha = h n_s /2$ is
the magnitude of the Magnus force ( $n_s$ is
the condensate density).
In terms of a frequency dependent viscosity, we find that
$\eta ( \omega ) \sim 0$ below the resonance, $\eta (
\omega_{res} ) \sim \rm{min} ( \eta_0 , \alpha )$, where 
$\eta_0$ is comparable to the Bardeen-Stephen value, and that
$\eta ( \omega ) \rightarrow 0$ as $\omega \rightarrow \infty$.

We have not attempted to average over a distribution of pinning constants.
To a first approximation, it is equivalent to an average over
frequencies, in a plot like the one shown in fig. (\ref{resl}).
The most noticeable effect will be the smoothing of the 
resonance shown there. We have also not analyzed the interaction
between vortices pinned in different regimes. The voltages induced
by the  vortex motion tend to be higher for those vortices close to
to resonance, and it is likely that they will
entrain the others. Finally, we do not consider here additional
sources of dissipation, like the existence of subgap states in
a d-wave superconductor\cite{Parks}. These effects will be analyzed
elsewhere.

The experiments reported so far focus mainly on the temperature dependence
of the vortex viscosity and pinning constant\cite{GTD,Parks}.
Pulse experiments, like those presented in\cite{Parks}, may be difficult
to interpret, as the exponential suppression of the vortex viscosity
at low velocities implies the existence of
long lived transients (which may be observed, however, as
echo signals). It would be interesting 
if the non linear effects discussed here can be confirmed experimentally.

\begin{figure}
\caption{
Trajectories followed by vortices driven at different frequencies.
a) $\omega = 0.05$. b) $\omega = 0.15$. c) $\omega = 0.5$. d) $\omega
= 5$. 
The ^^ ^^ ideal " resonance frequency described in the text is,
in these units, 
$\omega_{res} = k_p / ( 2 \pi \alpha ) = 0.159$. For typical
experimental parameters, $\omega_{res} \sim 2 \times 10^{11}$ Hz.
As mentioned in the
text, a reasonable unit of length is $\sim 40$\AA .}
\label{vortex}
\end{figure}

\begin{figure}
\caption{
Time dependence of the vortex velocities for the four cases
shown in fig. 1. The unit of velocity is $v_0 \sim 4$ Km/s.
Full line, $v_y$. Dashed line $v_x$. $T$ is
the period.} 
\label{vel}
\end{figure}

\begin{figure}
\caption{
Frequency dependence of the vortex resistivities as function of
frequency. The lowest harmonic of the resistivity is shown here.
Full line, Re $\rho$. Dotted line, Im $\rho$. Resistivity units are
arbitrary.}
\label{resl}
\end{figure}

\begin{figure}
\caption{Second (upper panels) and third (lower panels) harmonics of
the resistivity. The conventions are as in fig.3. The units are
the same as the ones used in fig. 3.}
\label{resh}
\end{figure}
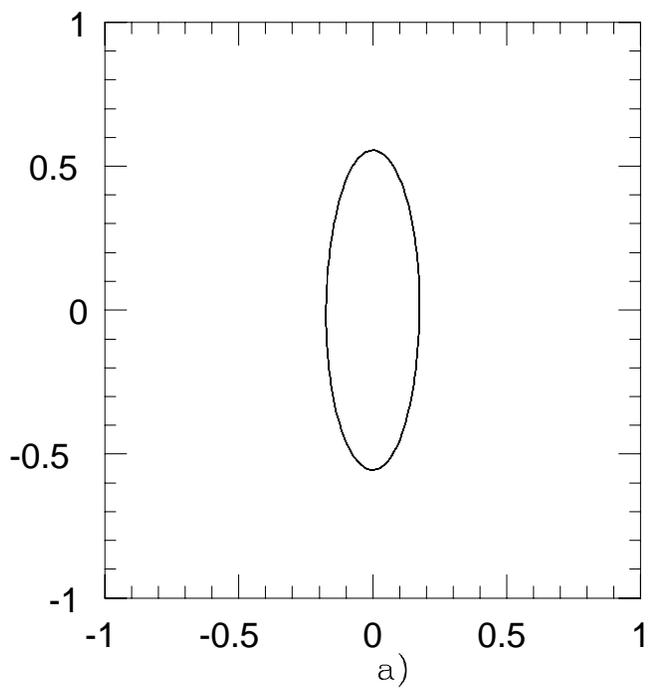
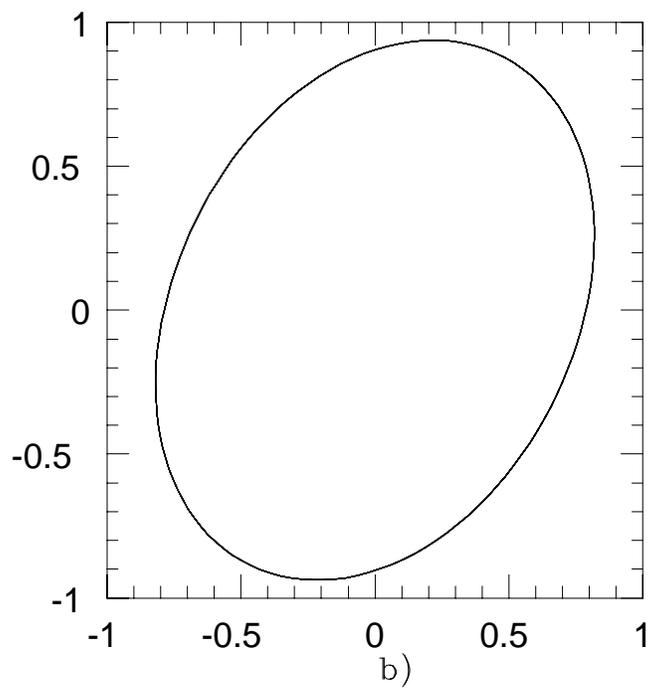
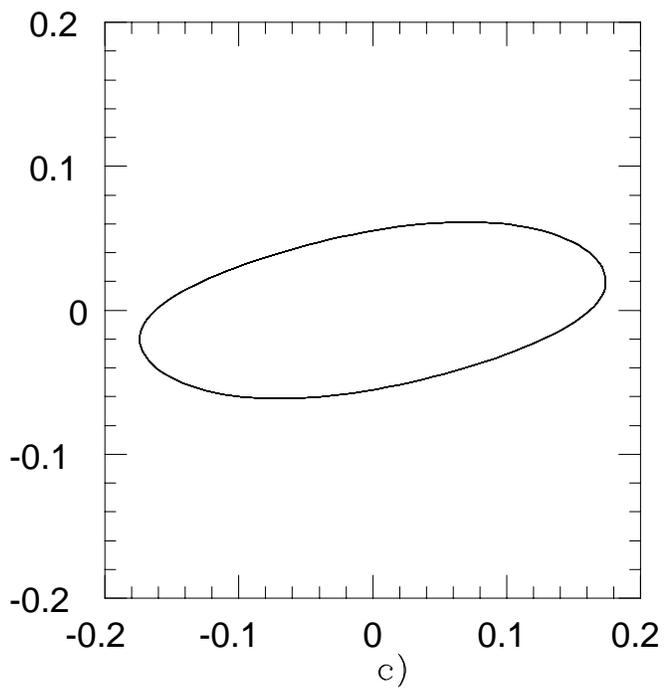
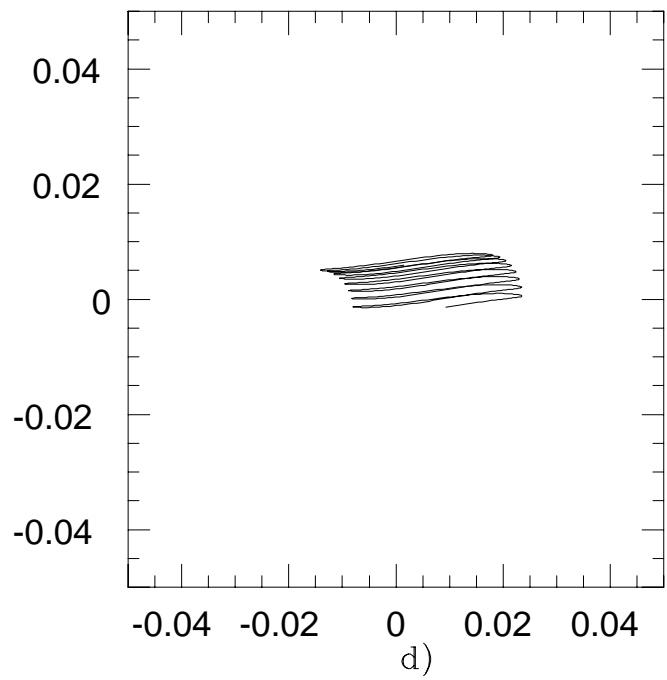
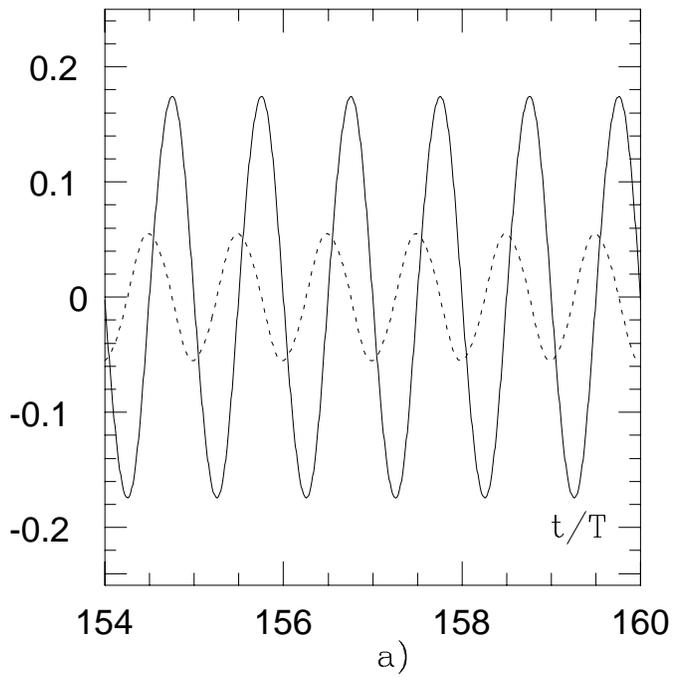
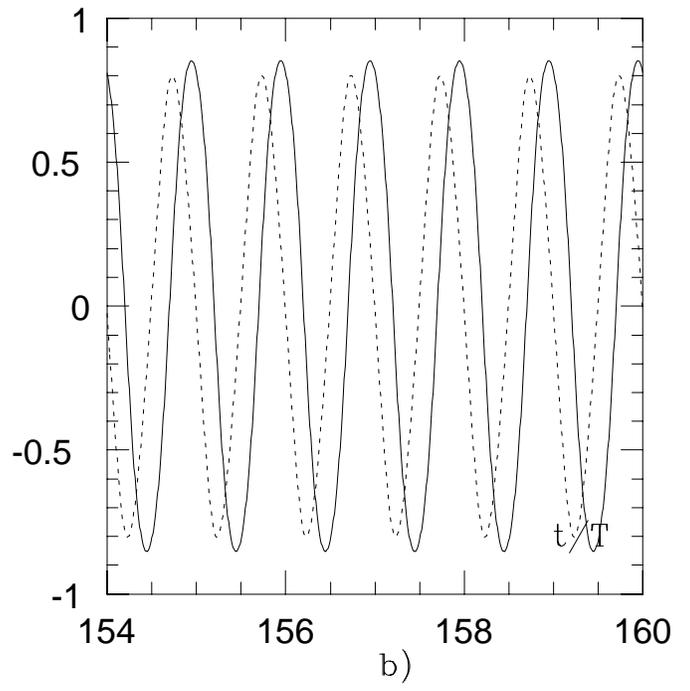
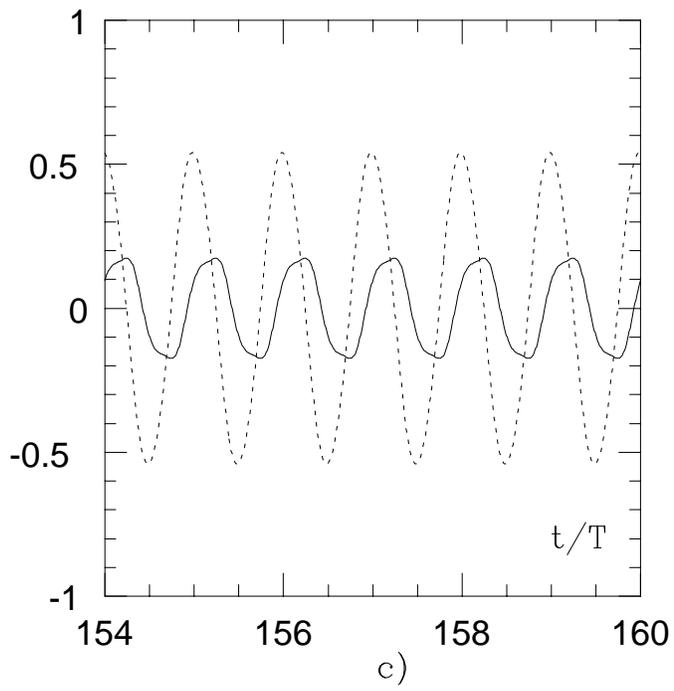
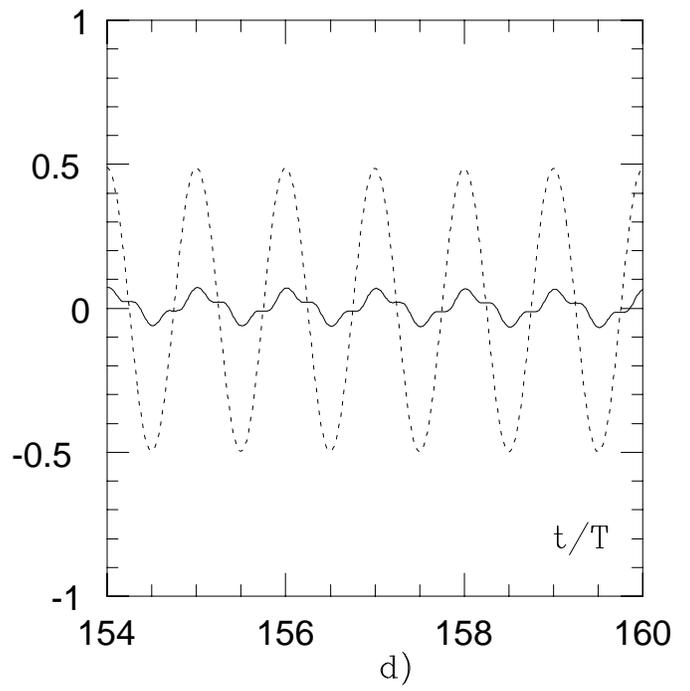
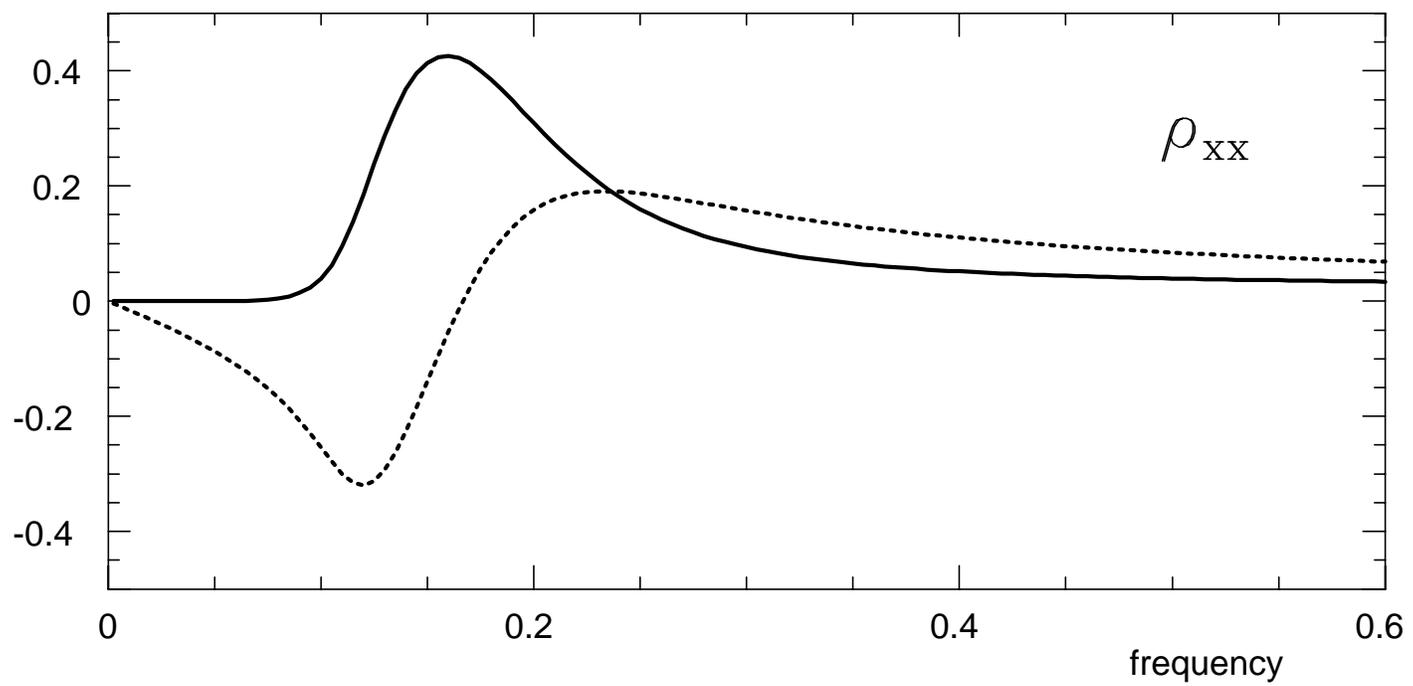
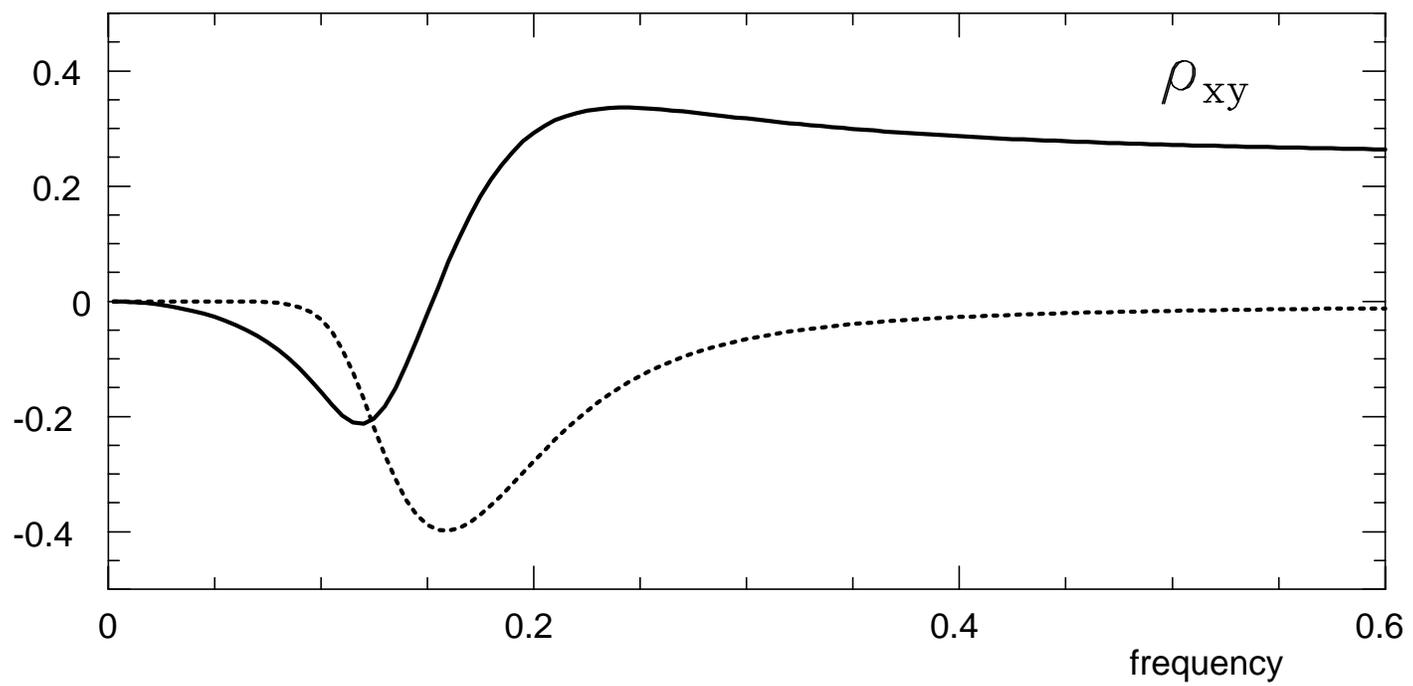
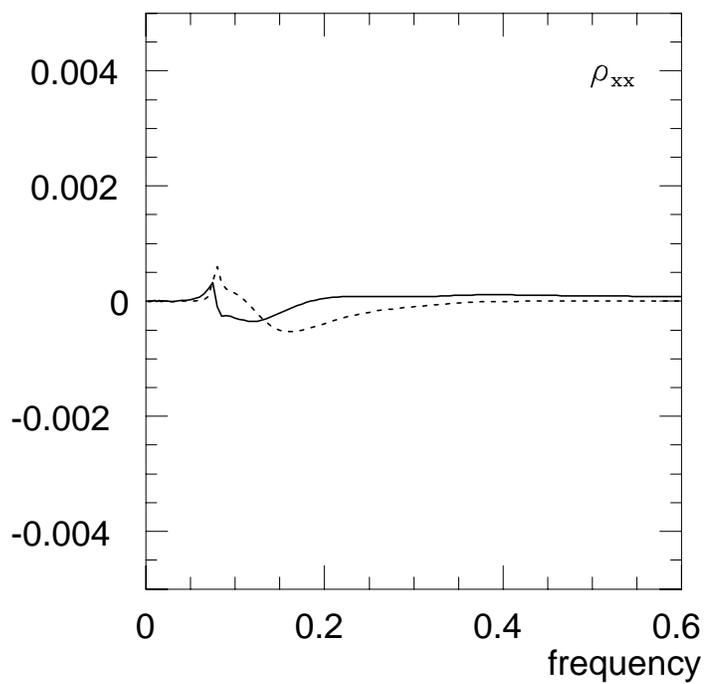
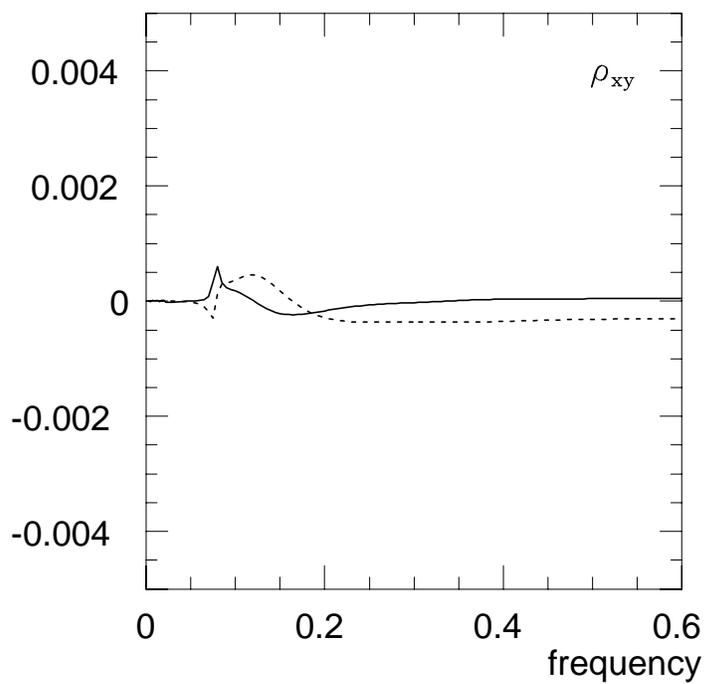
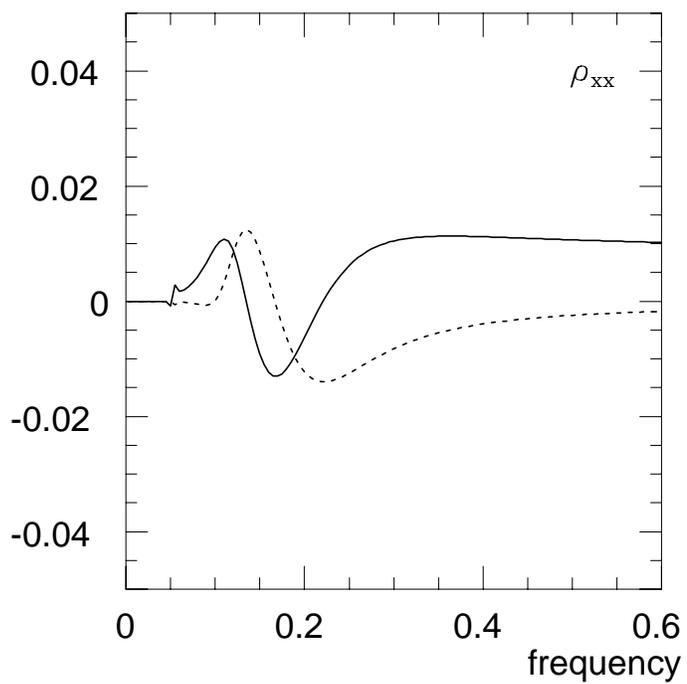
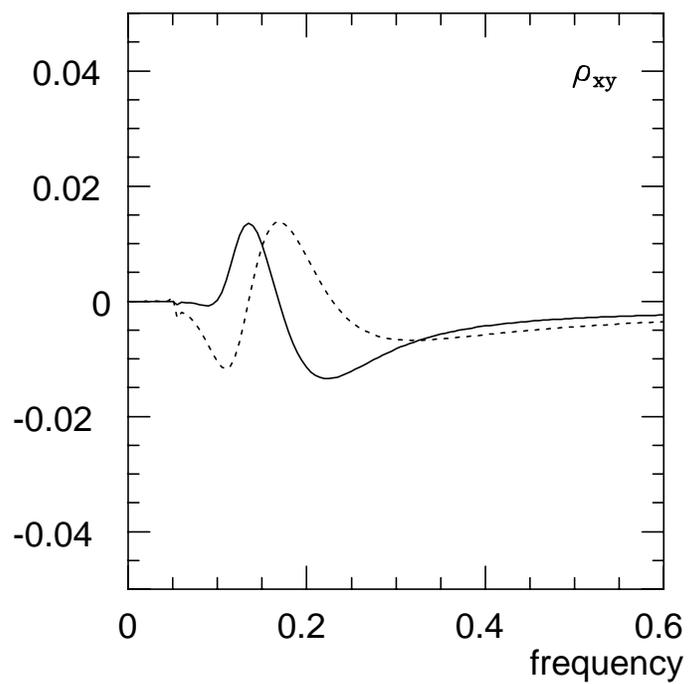
\end{document}